\documentstyle[aps,twocolumn]{revtex}
\begin{document}
\draft

\title{Optical pumping of quantum dot nuclear spins}

\author{A. Imamo\=glu$^1$, E.~Knill$^2$, L.~Tian$^3$, and P.~Zoller$^3$}

\address{$^1$ Institute of Quantum Electronics,
ETH H\"onggerberg HPT G12, CH-8093 Z\"urich, Switzerland}

\address{$^2$ Los Alamos National Laboratory, MS B265, Los Alamos, NM 87545, USA}

\address{$^3$ Institut f\"ur Theoretische Physik, Universit\"at Innsbruck,
Technikerstrasse 25, A-6020 Innsbruck, Austria}

\date{\today}

\maketitle

\begin{abstract}
An all-optical scheme to polarize nuclear spins in a single
quantum dot is analyzed. The hyperfine interaction with randomly
oriented nuclear spins presents a fundamental limit for electron
spin coherence in a quantum dot; by cooling the nuclear spins,
this decoherence mechanism could be suppressed. The proposed
scheme is inspired by laser cooling methods of atomic physics and
implements a "controlled Overhauser effect" in a zero-dimensional
structure.
\end{abstract}


\draft

\vskip2pc \narrowtext

One of the principal features that distinguish a quantum dot (QD)
from an atom is the completely different role that hyperfine
interactions play in the two systems. In contrast to valence
electrons of an atom, a quantum dot (conduction band) electron has
confinement length-scales that extend over many lattice sites. As
a result, a single electron spin interacts with $N \simeq 10^3 -
10^5$ nuclei and the interaction strength with each nucleus is
reduced by a factor determined by the probability of finding the
electron at that lattice site ($\sim 1/N$). Equivalently, the
electron hyperfine field seen by each nucleus will be extremely
weak and the nuclei will remain unpolarized. This random nuclear
spin orientation presents a fundamental decoherence mechanism for
an electron spin confined in a quantum dot. Recently, Khaetskii
{\sl et al.} \cite{Loss02} and Merkulov {\sl et al.}
\cite{Efros02} have analyzed this decoherence mechanism.

In this Letter, we propose an all-optical technique to polarize
the nuclear spins interacting with a single quantum dot electron
spin. The basic idea is to use the hyperfine coupling to induce a
controlled electron-nuclear spin-flip process. This can be
achieved by changing the energy of the initial (spin-up)
electronic state using the ac-Stark effect \cite{ac-stark}, in
order to allow for resonant electron-nuclear spin flip to take
place. When the spin-flip is completed, the electron spin is
re-flipped into its original state using a laser induced
$\pi$-pulse followed by spontaneous emission. Starting from a
random unpolarized ensemble nuclear spin state, each step as
described above flips one nuclear spin, albeit in a collective
way. Resonance between the spin-up and down electronic states can
also be achieved in sub-nanosecond time-scales using the
electric-field dependence of the electron g-factor \cite{Salis01},
without requiring a laser induced ac-Stark effect.

Before proceeding, we note that partial polarization of nuclear
spins using hyperfine interactions and optical pumping is well
studied \cite{Overhauser}. More recently, dynamical polarization
of lattice nuclei (the Overhauser effect) has been demonstrated in
interface QDs which form due to monolayer fluctuations of the
thickness of a GaAs quantum well \cite{Gammon}: here a circularly
polarized laser creates electrons in a well-defined spin-state,
which in turn polarizes the nuclear spins via hyperfine contact
interaction. Spin-flipped electrons are then removed from the
system by radiative recombination, maintaining a relatively high
degree of spin polarization for electrons and the nuclei
\cite{Gammon}. One limitation of this dynamic polarization scheme
is the fact that in a QD, creation of nuclear spin polarization
eventually makes joint electron-nuclear spin-flip processes
energetically forbidden due to the large electron Zeeman energy
induced by the nuclear magnetic field. In addition, optical
Overhauser effect relies on fast hole-spin relaxation for the
removal of the spin-flipped electron by radiative recombination.
However, recent experiments \cite{Marie01} demonstrate that
hole-spin relaxation is significantly slower in small quantum
dots. It is essential to overcome these two difficulties in order
to achieve a high level of nuclear spin polarization in QDs.

We consider a quantum dot where a single conduction band electron
interacts with $N \simeq 10^4$ (spin 1/2) nuclei. We assume that
the interaction with $i^{th}$ nucleus is proportional to the
absolute value of the electron wave-function squared at that site
($\alpha_i$, with $\sum_i \alpha_i = N$). The Hamiltonian
describing this interaction is the hyperfine contact interaction
\cite{Loss02}
\begin{eqnarray}
\hat{H}_{int} & = & \frac{A}{N} \sum_i \alpha_i \left[ \frac{1}{2} \hat{I}_z^i
\hat{\sigma}_z + \hat{I}_+^i \hat{\sigma}_- + \hat{I}_{-}^i \hat{\sigma}_{+}
\right], \label{eq1}
\end{eqnarray}
where $\hat{I}_k^i$ and $\hat{\sigma}_k$denote the Pauli operators
for the $i^{th}$ nucleus and the electron, respectively.
$\hat{\sigma}_+ = |\uparrow \rangle \langle \downarrow|$ is the
electron spin-flip operator. A is an effective hyperfine
interaction constant that takes into account the coupling of all
the nuclei in the unit cell; Merkulov et al. estimate A to be $90
\mu{\rm eV}$ for GaAs \cite{Efros02}.

We assume that the QD is subject to a large constant magnetic
field that removes the degeneracy of the spin-up and down states;
for an electron g-factor $g_e \sim 2$, we expect an energy
difference of 1 meV with B = 5 Tesla. The magnetic field is
applied along the direction of strong confinement; i.e. $\hat{z}$
for QDs grown by molecular beam epitaxy. For temperatures ($T \sim
3 K$) typical of magneto-optical cryostats, the electron is
spin-polarized (in the spin-up state) and the nuclei are not:
\begin{eqnarray}
| \Psi \rangle & = & | \psi\rangle_e \otimes | \psi\rangle_N \, = \,
\hat{\sigma}_{+} \prod_j^M \hat{I}_{+}^j |\phi \rangle , \label{eq2}
\end{eqnarray}
where the product of nuclear spin operators is over a random set
of nuclei. For unpolarized nuclei $M \sim N/2$. $|\phi \rangle \,
= \, \hat{e}_{\downarrow}^{\dagger} \prod_{i=1}^N
\hat{n}_{i,\downarrow}^{\dagger} | 0 \rangle$, where
$\hat{e}_{\downarrow}^{\dagger}$ and
$\hat{n}_{i,\downarrow}^{\dagger}$ correspond to the creation
operator of the spin-down electron and the $i^{th}$ nucleus,
respectively.

The electron spin dynamics in a QD takes place on timescales much
shorter than that of the nuclear spin system. As a result, we can
assume that the nuclear spins are in a random but constant state
during the timescale over which the electron spin is manipulated.
The effective (nuclear) magnetic field \cite{Loss02,Efros02} seen
by the QD electron is $B_z^{eff} \, \sim \, A / (\sqrt{N} g_e
\mu_B)$, where $\mu_B$ is the Bohr magneton. The magnetic field
along $x$ and $y$ directions have the same expectation value for
unpolarized nuclei.

Interactions with a classical time-dependent laser field are governed by the
Hamiltonian
\begin{eqnarray}
\hat{H}_{laser} & = & \hbar  \left[ \Omega_+(t) (\hat{e}_{\downarrow}
\hat{h}_{3/2} + \eta \hat{e}_{\uparrow} \hat{h}_{3/2}) e^{i \Delta t} +
c.c. \right] \, \nonumber \\
&+& \, \hbar \left[ \Omega_-(t) (\hat{e}_{\uparrow} \hat{h}_{-3/2} + \eta
\hat{e}_{\downarrow} \hat{h}_{-3/2}) e^{i \Delta t} + c.c. \right], \label{eq3.5}
\end{eqnarray}
where $\hat{h}_{\pm 3/2}$ denotes a valence band hole state with
angular momentum projection $j_z=\pm 3/2$. $\Omega_{+}(t)$
($\Omega_{-}(t)$) is the time-dependent Rabi frequency of the
right (left) hand circularly polarized laser field propagating
along the $z-$direction and interacting with the strongly allowed
QD transition that satisfies the $j_z$  selection rules. Due to
heavy-light hole mixing of the valence band states, these
selection rules are relaxed in actual QD structures, leading to
non-zero but small coupling ($\eta \ll 1$) to optical fields that
violate the $j_z$ selection rules \cite{forbidden}. The frequency
of the laser field determines the detuning $\Delta$ of the optical
transition.

In the presence of a large Zeeman splitting, electron-nuclear
spin-flip processes are forbidden by energy conservation
(Fig.~1(a)). The first step in the proposed scheme is the
application of a red-detuned left-hand circularly polarized (lcp)
laser pulse that creates a "spin-state dependent ac-stark effect"
that effectively cancels the Zeeman splitting of the electron
caused by the external magnetic field \cite{ac-stark}. While this
laser is on electron-nuclear spin flip process due to
$\hat{H}_{int}$ of Eq.(1) is energetically allowed (Fig.~1(b)).
The effective coupling coefficient for spin-flip process for a
random initial state is given by
\begin{eqnarray}
g_{spin-flip} & = & || \frac{A}{N} \sum_i \alpha_i \hat{I}_+^i \hat{\sigma}_- |
\Psi \rangle ||  , \label{eq4}
\end{eqnarray}
as a direct consequence of the collective enhancement of the
transition due to participation of many nuclei.  Since
$g_{spin-flip}$ is comparable to $g_e \mu_B B_{eff}$ (for
unpolarized nuclei), we expect significant probability for
spin-flip if we leave the laser on for $\tau \sim 1/g_{spin-flip}$
($\hbar = 1$). We estimate the spin-flip time for an electron in
an InAs QD to be less than 1 nsec \cite{Efros02}. Therefore
choosing a laser pulse-width of $\tau \sim 300 psec \, <
g_{spin-flip}^{-1}$ will yield a spin-flip probability $\sim 0.1$.
If an electron-nuclear spin flip event does take place, the
state-vector of the QD is $| \Psi \rangle_A \, = \,
 \hat{\sigma}_{-} \sum_i \alpha_i
\hat{I}_{+}^i | \Psi \rangle \, / \cal{N}$, where $\cal{N}$ ($\neq
N$ in general) is the normalization factor. Since collective
enhancement factor becomes smaller with increasing nuclear
polarization, it is desirable to increase $\tau$ as cooling
proceeds.

After the ac-stark laser is turned off, we turn on a resonant
right hand circularly polarized (rcp) laser field that realizes a
$\pi$-pulse on the quasi-forbidden electronic transition $
\hat{e}_{\downarrow}^{\dagger} |0 \rangle \, \rightarrow \,
\hat{e}_{\downarrow}^{\dagger} \hat{e}_{\uparrow}^{\dagger}
\hat{h}_{3/2}^{\dagger} |0 \rangle$ (with transition amplitude
$\propto \eta$), only if an initial electron-nuclear spin-flip
process has taken place due to the resonant hyperfine interaction
in the presence of the red-detuned lcp laser. If this is the case,
the excited trion state $ \hat{e}_{\downarrow}^{\dagger}
\hat{h}_{3/2}^{\dagger} |\psi\rangle_e$ is populated with
probability approaching unity. This excited state will relax down
predominantly to the electronic state $ |\psi\rangle_e =
\hat{e}_{\uparrow}^{\dagger} |0 \rangle$ by spontaneous emission
of a rcp photon with rate $\Gamma_{rad}$, thereby projecting the
electron spin onto the initial spin-up state (Fig.~1(c)). The
final state following spontaneous emission is
\begin{eqnarray}
| \Psi \rangle_C & = &  \frac{1}{\sqrt{\cal{N}}} \, \sum_i \alpha_i \hat{I}_{+}^i
| \Psi \rangle  . \label{eq8}
\end{eqnarray}

The successive application of two laser pulses followed by
spontaneous emission flips a single nuclear spin with probability
$\sim 0.1$ and constitutes the elementary step of the proposed
laser cooling scheme for nuclear spins. If electron-nuclear spin
flip due to hyperfine interaction does not take place, then the
applied $\pi$-pulse does not couple an occupied transition and the
whole system remains in its initial state $| \Psi\rangle$.
Residual coupling on the detuned strongly allowed transition
$\hat{e}_{\uparrow}^{\dagger} |0 \rangle \rightarrow
\hat{e}_{\downarrow}^{\dagger} \hat{e}_{\uparrow}^{\dagger}
\hat{h}_{3/2}^{\dagger} |0 \rangle$ by the rcp laser will result
predominantly in Rayleigh scattering down to the electronic state
$\hat{e}_{\uparrow}^{\dagger} |0 \rangle$. We note that the energy
of the spontaneously emitted photons (which indicate the presence
of a nuclear spin flip) and Rayleigh scattered photons (which
indicate unchanged nuclear spin state) differ by the electron
Zeeman splitting.

Having discussed the elementary step of near-deterministic
flipping of a single nuclear spin collectively, we next turn to
the question of the effectiveness of successive applications of
this cooling step in achieving large nuclear spin polarization.
First we note that $B_z^{eff}$ will change as the nuclear spin
polarization increases. For unpolarized nuclei $g_e \mu_B
B_z^{eff} \sim g_{spin-flip}$, whereas for nearly polarized nuclei
$g_e \mu_B B_z^{eff} \sim A \gg g_{spin-flip}$. Therefore, if the
magnitude of the ac-Stark shift simply cancels out the Zeeman
shift due to the external magnetic field, the spin-flip process
will become ineffective as the nuclear spin polarization
increases. By adjusting the intensity of the lcp laser, it should
be possible to change the magnitude of the ac-Stark shift to
ensure resonance condition for electron-nuclear spin-flip
processes, for all values of nuclear-spin polarization. Since each
spin-flip is accompanied by spontaneous emission of a photon, it
is in principle possible to estimate the degree of nuclear spin
polarization by counting the spontaneously emitted photons.
Alternatively, we can envision a laser pulse shape that will
ensure resonance condition for a sufficiently long interaction
time, for any $\langle \sum_i \hat{I}_z^i \rangle$.

The principal question that determines a limitation of the proposed scheme is the
probability of the nuclear-spin system evolving into a dark-state of the
Hamiltonian of Eq.(\ref{eq1}); i.e. if the nuclear-spin state after n-steps of
laser-induced collective spin-flip events ($|\psi\rangle_N^{(n)}$) satisfies
\begin{eqnarray}
\sum_i \alpha_i \hat{I}_+^i |\psi\rangle_N^{(n)} & = &  0  , \label{eq9}
\end{eqnarray}
then the prescribed procedure cannot be utilized to achieve further nuclear spin
polarization with the given $\hat{H}_{int}$. An illustrative example is the case
when $\alpha_i = 1, \forall i$ and the QD nuclear spin system is in the first
excited state with a single flipped nuclear spin. Of the $N$ states in this this
manifold, the only state with appreciable coupling ($g_{spin-flip} \sim
A/\sqrt{N}$) to the fully-polarized ground-state is the completely symmetric
state. The other $N-1$ asymmetric states satisfy Eq.~(\ref{eq9}).

In the limit of inhomogeneous electron-nuclear-spin coupling
($\alpha_i \neq \alpha_j, \forall i,j$) that is of practical
interest, total nuclear spin $\hat{I}_T^2$ is not conserved and
the limitation due to (quasi) dark states will only be relevant
for the sub-collection of nuclear spins for which $\alpha_k \simeq
\alpha_l$. A potential remedy in this case is provided by the fact
that the spatial wave-function of the electron confined in the QD
can be modified using external electric fields. This modification
will in turn alter the hyperfine interaction coefficients
$\alpha_i$. We can then use the feedback from the measurement of
emitted photons to change these coefficients as the cooling
progresses: if the number of detected photons falls below a
certain pre-determined level, this is a good indication that the
system has evolved into a quasi-dark state of $\hat{H}_{int}$ with
the current $\alpha_i$. Based on this information, we can
introduce an external electric field and increase its magnitude
(or change its orientation) until we increase the photon detection
rate. Alternatively, we can apply a random electric field with
coherence time shorter than the separation of successive
elementary cooling cycles to ensure that distant nuclei will have
$\alpha_i \neq \alpha_j$ for the majority of the elementary
cooling steps.

To evaluate the role of dark states in nuclear spin cooling, we
have carried out a numerical simulation of the proposed scheme for
a toy system consisting of 10 nuclei. When we choose a symmetric
Gaussian wave-function for the electron, we find that the nuclear
polarization saturates at $75 \%$ (Fig.~2 solid line). This
saturation is due to the dark (singlet) states of pairs of nuclei
with identical hyperfine coupling to the electron. We then shift
the electron wave-function by $0.5 a_L$, where $a_L$ is the
lattice constant. Since the new hyperfine coupling distribution
has a different set of dark states, the polarization increases
abruptly and then saturates at a higher level. Further small
shifts of the electron wave-function results in $95\%$
polarization of the nuclear spins (Fig.~2). In contrast, for a
fixed (Gaussian) $\alpha_i$ with $ \alpha_i  \neq \alpha_j,
\forall i \neq j$, the nuclear system reaches $ > \%99$ spin
polarization in much shorter time-scales.

For actual QDs, we will have $\alpha_i \sim \alpha_j$ for nuclei
that are nearest-neighbors, even when a varying external electric
field is applied. This will in turn result in the slowing down of
the cooling process. A possible remedy for moving neighboring
nuclei out of quasi-dark states is provided by the first term in
$\hat{H}_{int}$ of Eq.~(\ref{eq1}) which acts to randomize the
relative phase between product states with identical $\langle
\hat{I}_z \rangle = \sum_i \langle \hat{I}_z^i \rangle$ that make
up a dark state. For $\tau >> 10^{-5}$~sec, we estimate that two
product states for which only two neighboring nuclei have
differing $I_z$ values, can accumulate phases that differ by
$\pi$. Therefore, we expect the nuclear system to move out of a
dark state in less than $100 \, \mu$~sec and the laser cooling to
proceed.

The achievable nuclear spin temperature is limited by the nuclear
spin diffusion from the (highly polarized) QD nuclei to the
(partially-polarized) nuclei of the surrounding semiconductor. The
physical mechanism for nuclear spin diffusion could be provided by
the (secular) terms in nuclear dipole-dipole interactions which
allow for resonant spin exchange between two nuclei while
preserving $\langle \hat{I}_z \rangle$. For QDs embedded in a
semiconductor of a different type, we expect the different
g-factors for the nuclei in the two semiconductors to largely
inhibit spin diffusion into the surrounding material. This should
be the case for CdSe/ZnS core-shell nanocrystals and InAs
self-assembled QDs. For such QDs, nuclear dipole-dipole
interactions with typical time-scales $\tau_{nuclear} \sim
10^{-4}$ sec will act to help the nuclear spin cooling by
transferring the polarization to those QD nuclei which have small
wave-function overlap ($\alpha_i \ll 1$) with the QD electron. In
addition, nuclear spin-flips due to resonant dipole-dipole
interaction will also be effective in moving the total QD nuclei
system out of dark states.

For electrically defined structures, the semiconductors that make up the QD and
(part of) the barrier are identical. In this case, nuclear dipole-dipole
interactions can cause spin diffusion into the barrier and limit the effectiveness
of laser cooling \cite{gated-dots}. A possible remedy for nuclear spin cooling in
such QDs can be obtained from NMR techniques, such as magic angle time-dependent
fields, that can be used to eliminate dipolar interactions to a large extent
\cite{NMR}. We also note that recent experiments on electrically defined QDs
showed a spin diffusion time of $800$~sec - more than 6 orders of magnitude longer
than the typical timescale for dipolar interactions \cite{Tarucha03}.

To make a worst case estimate of the spin cooling time we can
assume that the QD nuclear spin system moves into a dark state
after each electron-nuclear spin flip event. Since we estimate the
time to move out of a dark state to be $10^{-4}$~sec, the cooling
time for a system of $N=10^4$ nuclei would be $\sim 1$~sec. For
nuclear spin diffusion (i.e. $T_1$) times exceeding $100$~sec,
$\ge 99\%$ polarization of the nuclear spins could be possible
\cite{laser-heat}.

In summary, we have described an all-optical method that flips the
nuclear spins in a pre-determined direction. Successive
application of the spin-flip procedure will realize laser cooling
of nuclear spins in a zero-dimensional structure. The elementary
step of near-deterministic nuclear spin flip process can be used
to generate highly entangled states of the nuclei, even before
significant nuclear spin polarization is achieved. It has been
shown recently that such states can have completely different
signatures for electron spin dynamics, as compared to unpolarized
nuclei in a product state \cite{Schliemann02}. Another promising
application of fully polarized nuclear spins is in quantum state
storage of an electron spin state in collective excitations of
nuclear spins \cite{Lukin03}.

The Authors acknowledge useful discussions with L.~M.~Duan,
Al.~L.~Efros, D.~Loss, and G.~Salis.

\begin{figure} \caption{All-optical manipulation of electron-nuclear spin-flip
in a single quantum dot. (a) In the presence of a large Zeeman splitting,
electron-nuclear spin-flip events are energetically forbidden. (b) Introduction of
a red-detuned laser field can effectively cancel the Zeeman splitting and allow
for resonant spin-flip processes due to hyperfine interaction. (c) The
spin-flipped electron is re-pumped into the initial state by a combination of a
$\pi$-pulse, followed by spontaneous emission.} \label{fig1}
\end{figure}

\begin{figure} \caption{Optical pumping of a quantum dot with 10 nuclear spins.
We assume a Gaussian wave-function for the electron and choose its
center such that pairs of nuclear spins have identical $\alpha$.
Starting from a completely unpolarized nuclear-state with density
matrix $\rho_{i} = 1/2^N$, we find that the nuclear polarization
saturates at $0.75$ (solid line). This saturation is a clear
indication of the influence of dark states. We then shift the
electron wave-function (after $5000$ pulse sequences) in a way to
ensure that $\alpha$'s are identical for different sets of nuclei:
since the new coupling distribution has a different set of dark
states, the polarization increases abruptly and then saturates at
a higher level. Further shifts after $15000$ and $20000$ pulse
sequences results in $95\%$ polarization of the nuclear spins
(solid line). The dashed line shows $\langle \hat{I}_z \rangle$
when the electron wave-function is shifted by the same small
amounts, but now after every $50$ cooling pulse sequences. The
dotted line shows the simulation result for a single Gaussian
wave-function that ensures $\alpha_i \neq \alpha_j, \forall i \neq
j$: in this case, there are no dark states and the final nuclear
spin polarization $\ge 99 \%$ is achieved after only $2000$ pulses
(inset).} \label{fig2}
\end{figure}


\begin{references}
\bibitem{gated-dots} Application of the proposed technique to electrically defined
quantum dots is only possible if the electron-hole attraction overcomes the
repulsive potential for holes and results in a discrete optical excitation
spectra.
\bibitem{laser-heat} For $\eta = 0.01$,
$\Gamma_{rad} =  10^9 s^{-1}$ and an electron Zeeman splitting of
$1 {\rm meV}$, the laser heating rate of the nuclear spins is
about $10^4$ times smaller than the cooling rate.
\end{references}
\end{document}